\documentclass[final,1p,times]{elsarticle} 

\usepackage{graphicx}
\usepackage{amssymb} 
\usepackage{amsthm} 
\usepackage{lineno}

\journal{Nuclear Physics A} 

\begin{document}

\begin{frontmatter} 

\title{Study of pseudorapidity dependence of the anisotropic flow with ALICE at the LHC}

\author{Alexander Hansen (for the ALICE\fnref{col1} Collaboration)}
\fntext[col1] {A list of members of the ALICE Collaboration and acknowledgements can be found at the end of this issue.}
\address{Niels Bohr Institute, Blegdamsvej 17, 2100 Copenhagen, Denmark}

\begin{abstract} 
We report on the pseudo-rapidity dependence of the charged particle anisotropic flow in Pb-Pb collisions at $2.76$ TeV. The measurement is done over a wide range of pseudo-rapidity, $|\eta|<5$ using the forward detectors of ALICE at the LHC. 

Results are obtained from two- and multi-particle correlation techniques with the latter being less sensitive to non-flow effects. Elliptic flow longitudinal scaling, comparison with RHIC data and AMPT model calculations for the LHC are discussed.
\end{abstract} 

\end{frontmatter} 


\section{Introduction}
\label{s:Introduction}
In ultra-relativistic heavy-ion collisions a new state of matter known as Quark-Gluon Plasma (QGP) is produced. A key observable in the study of QGP is azimuthal anisotropy in particle production wrt. the collision symmetry plane, $\Psi_n$ \cite{Ollitrault:1992bk, Voloshin:2008dg}. The anisotropies are described by coefficients, $v_n$, in a Fourier decomposition of the azimuthal yields with respect to the corresponding $\Psi_n$. Anisotropic flow harmonics are calculated as an average over all particles, $v_n = \left \langle \cos \left [ n (\varphi - \Psi_n) \right ] \right \rangle$, where $\varphi$ is the azimuthal angles of the particles.

For many years only the first and second symmetry plane were considered to be of importance. Recent developments have shown that higher harmonics are present and provide important information about the QGP \cite{Alver:2010gr}. Nowadays the flow harmonics $v_1-v_6$ are all being reported on \cite{ALICE:2011ab,ATLAS:2012at}, and can help set tighter limits on the shear viscosity, $\eta/s$, of the QGP.

The elliptic flow coefficient, $v_2$, has previously been measured over a wide pseudorapidity-range, $\eta$, in Au-Au collisions over about an order of magnitude in collision energy ($\sqrt{s_{\rm NN}} = 19.6 - 200$ GeV) \cite{Back:2004zg}. $v_2$ was found to be independent of collision energy when observed in the rest frame of one of the colliding nuclei, an effect known as longitudinal scaling. 

Here we report on elliptic flow ($v_2$) and triangular flow ($v_3$) as measured over a wide pseudorapidity range, $-3.75 < \eta < 5$, in Pb-Pb collisions at $\sqrt{s_{\rm NN}} = 2.76$ TeV and compare to models and previous measurements at RHIC. Elliptic flow fluctuations has recently been found to be independent of $p_{\rm T}$ up to very high transverse momenta \cite{Abelev:2012di}, and here we report on the elliptic flow fluctuations vs. rapidity.

\section{Analysis details}
\label{s:Analysis}
The ALICE \cite{Aamodt:2008zz} minimum bias trigger was used for the event selection. Only events with a valid centrality estimate and primary vertex $|v_z| < 10$ cm were accepted. In total about $10.6$ million events from the 2010 data taking period were analyzed. The centrality is estimated using the VZERO detector, a pair of scintillator arrays covering $-3.7 < \eta < -1.7$ and $2.8 < \eta < 5.1$ in pseudorapidity.

The flow analysis is done on charged particle hits in the Forward Multiplicity Detector, a silicon strip detector which covers $-3.4 < \eta < -1.7$ and $1.7 < \eta < 5$. At mid-rapidity clusters from the innermost layer of the Silicon Pixel Detector, the inner part of the ALICE Inner Tracking System, are used. Using hits and clusters for the analysis, allows to measure flow down to almost zero $p_{\rm T}$. 

The analysis is based on analytical calculations for two- and four-particle cumulants \cite{Bilandzic:2010jr}, written as $v_n\{2\}$ and $v_n\{4\}$ respectively. The two-particle cumulant is enhanced by flow fluctuations, $\sigma_{v_n}$, and non-flow, $\delta_n$, such that $v^2_n\{2\} = \langle v_n \rangle^2 + \sigma^2_{v_n} + \delta_n$ and the four-particle cumulant has a negative contribution from flow fluctuations and is unaffected by non-flow, $v^2_n\{4\} = \langle v_n \rangle^2 - \sigma^2_{v_n}$. For this analysis the contribution by non-flow is subtracted using azimuthal correlations, $v_n\{2\}^{pp}$, from pp collisions at $\sqrt{s}=2.76$ TeV: $\delta_n^{cent}=v^2_n\{2\}^{pp}\cdot\frac{M^{pp}}{M^{cent}}$, where $M$ is the multiplicity. The different sensitivity to flow fluctuations can be used to estimate the flow fluctuations as \cite{Voloshin:2008dg}:
\begin{equation}
\frac{\sigma_{v_n}}{\langle v_n \rangle} \approx \sqrt{\frac{v^2_n\{2\}-v^2_n\{4\}}{v^2_n\{2\}+v^2_n\{4\}}}
\end{equation}

\begin{figure}[ht]
  \hspace{-1.3cm}
  \begin{minipage}[b]{0.55\linewidth}
    \centering
    \includegraphics[width=1.1\textwidth]{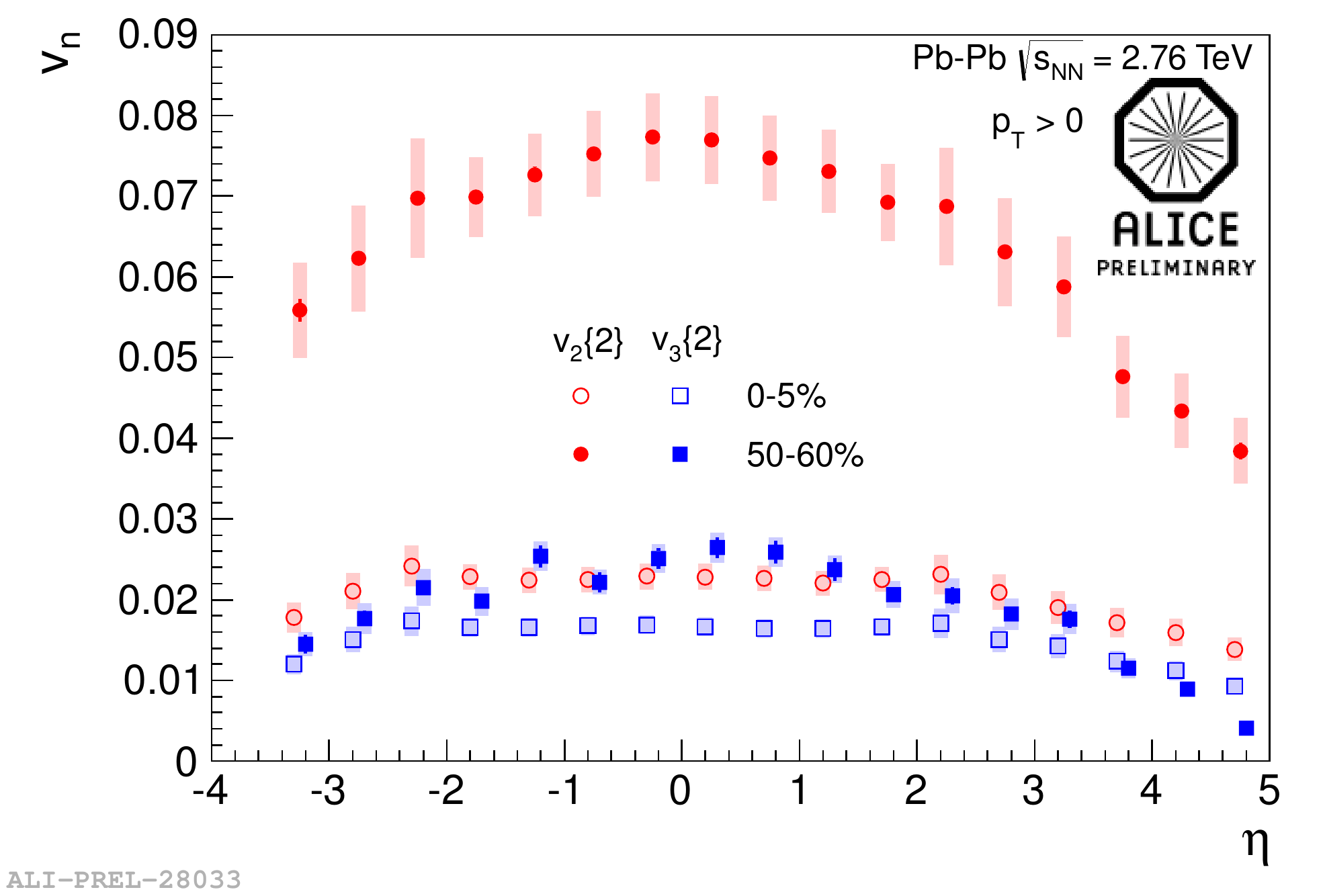}
    \caption{(color online) $v_2\{2\}$ and $v_3\{2\}$ vs. $\eta$ for very central events $(0-5\%)$ and more peripheral $(50-60\%)$.}
    \label{f.flow1a}
  \end{minipage}
  \hspace{0.40cm}
  \begin{minipage}[b]{0.55\linewidth}
    \centering
    \includegraphics[width=1.1\textwidth]{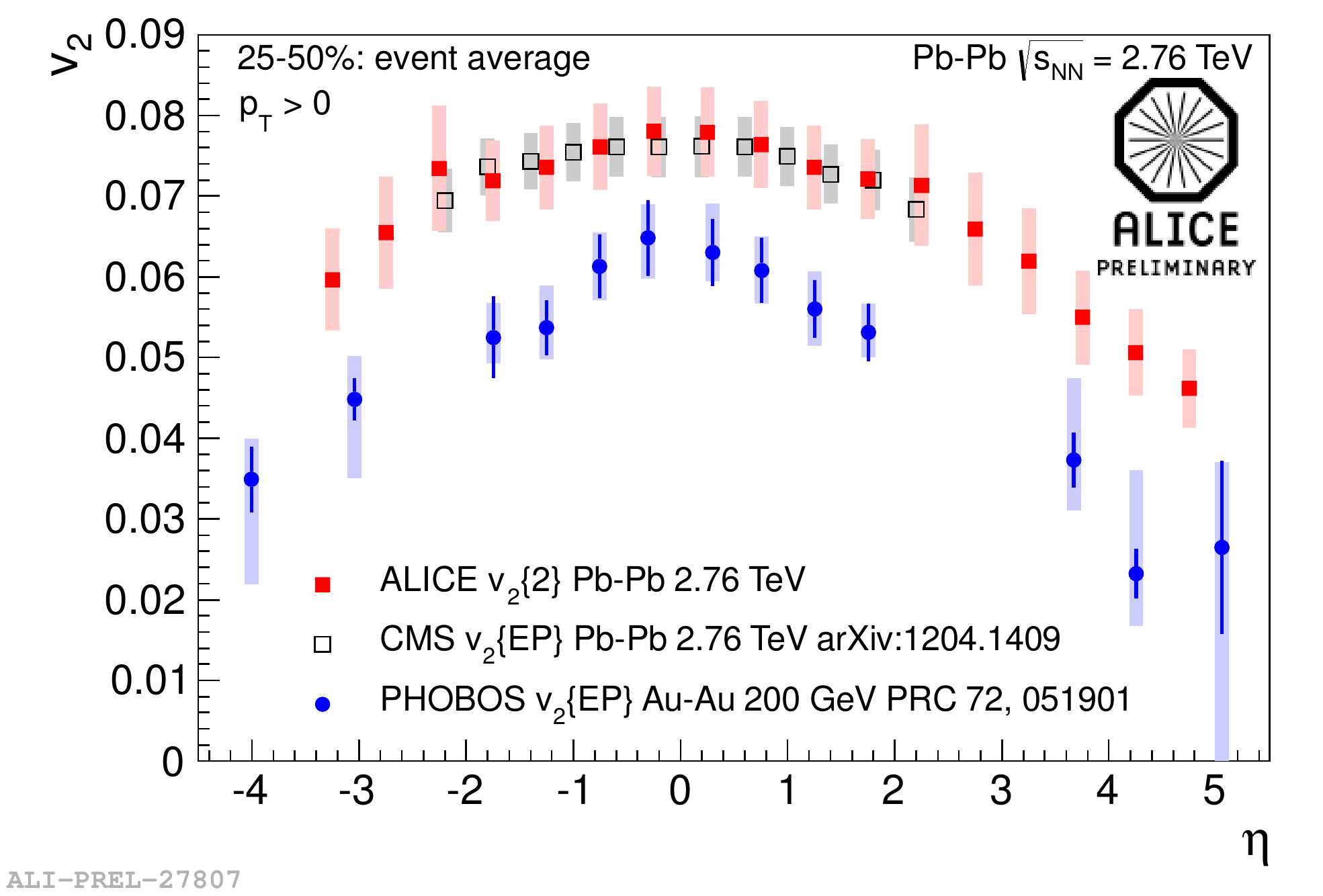}
    \caption{(color online) $v_2$ comparison with PHOBOS\cite{Back:2004mh} and CMS\cite{Chatrchyan:2012ta} for $25-50\%$ central events.}
    \label{f.flow1b}
  \end{minipage}
\end{figure}
\section{Results}
\label{s:Results}
Results from two-particle cumulant calculations of elliptic and triangular flow for very central and more peripheral events (Fig.~\ref{f.flow1a}) clearly show that $v_2$ has a strong centrality dependence for all rapidities, while $v_3$ has a weak centrality dependence. 
It has previously been shown that there is a $20-30\%$ increase in $v_2$ going from the RHIC top energy to LHC at mid-rapidity \cite{Aamodt:2010pa}, a comparison over a wide rapidity range (Fig.~\ref{f.flow1b}) shows that at forward-rapidities the increase can be larger than $30\%$. 

\begin{figure}[ht]
  \begin{minipage}[b]{0.55\linewidth}
  \hspace{-1.6cm}
    \centering
    \includegraphics[width=1.1\textwidth]{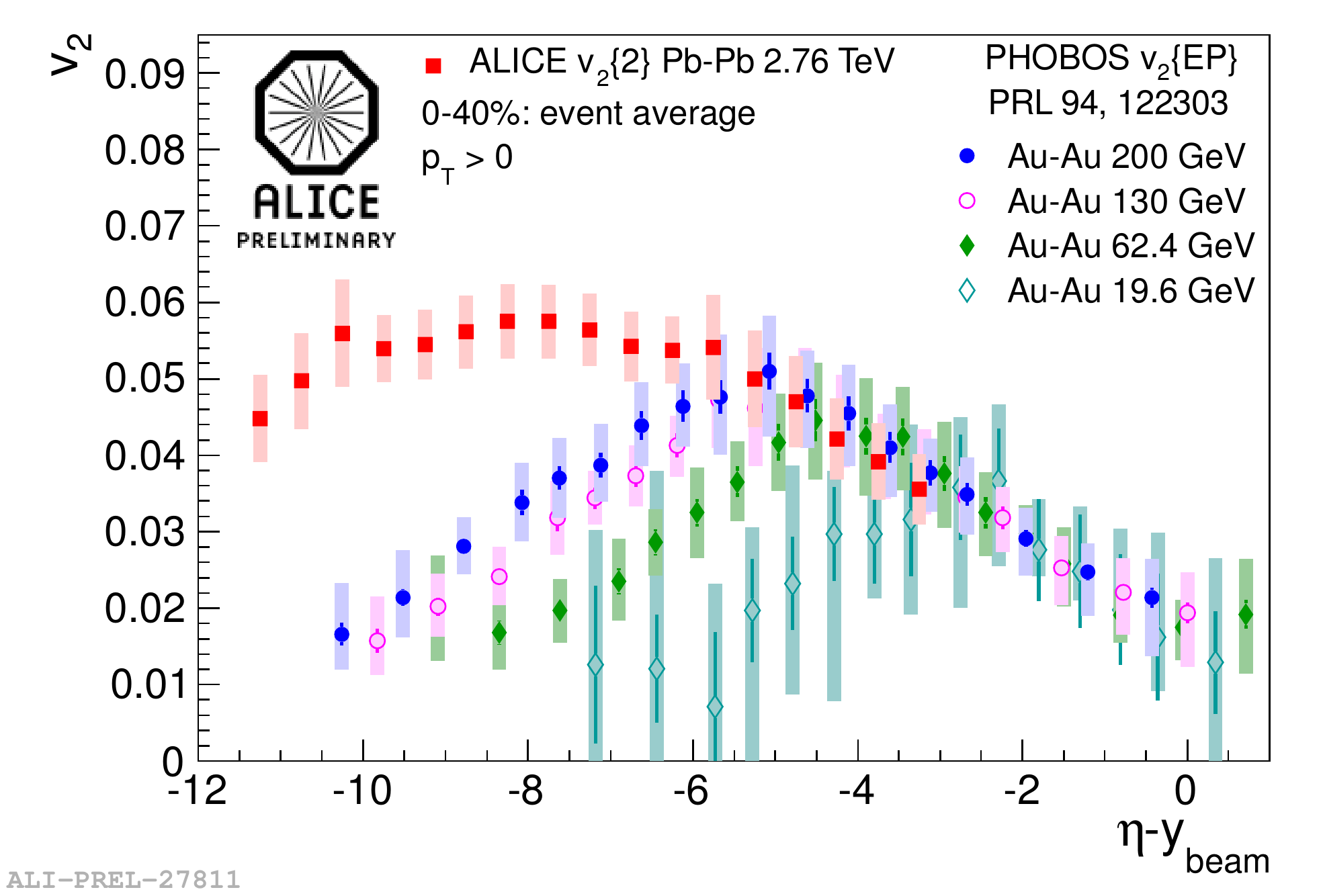}
    \caption{(color online) longitudinal scaling of $v_2$ over two orders of magnitude in collision energy, with data from PHOBOS\cite{Back:2004zg}.}
    \label{f.flow2a}
  \end{minipage}
  \hspace{0.30cm}
  \begin{minipage}[b]{0.55\linewidth}
  \hspace{-1.6cm}
    \centering
    \includegraphics[width=1.1\textwidth]{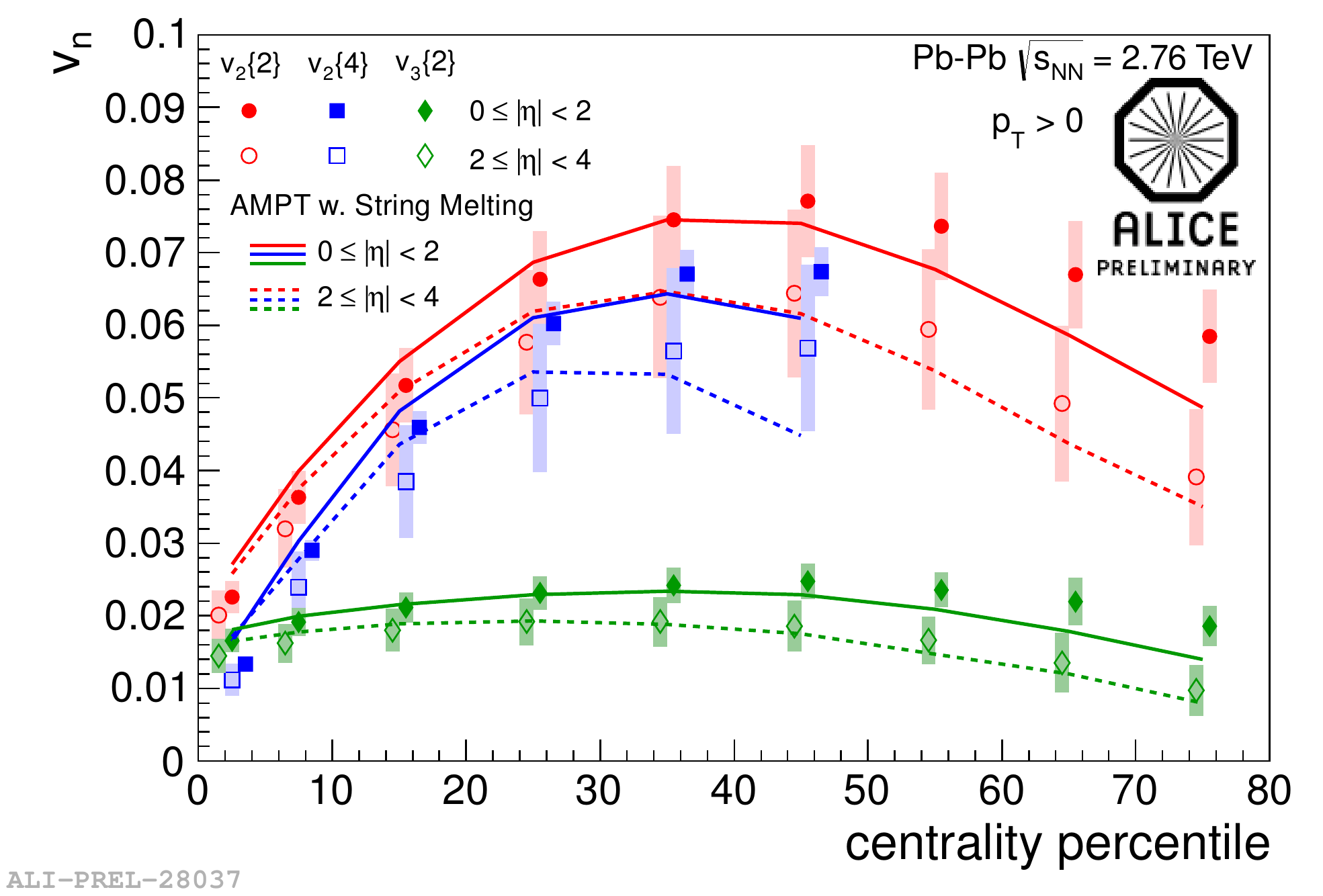}
    \caption{(color online) $v_2$ and $v_3$ vs. centrality compared to AMPT with parameters tuned to LHC mid-rapidity results.}
    \label{f.flow2b}
  \end{minipage}
\end{figure}

$v_2$ plotted as a function of pseudorapidity measured from a beam rapidity (Fig.~\ref{f.flow2a}) exhibits longitudinal scaling previously observed at RHIC \cite{Back:2004zg} and for directed flow at the LHC \cite{Selyuzhenkov:2011zj}. 
The AMPT model \cite{Lin:2004en} (Fig.~\ref{f.flow2b}) with parameters tuned to mid-rapidity LHC results for semi-central events \cite{Xu:2011fi} gives a good description of $v_2\{2\}$, $v_2\{4\}$ and $v_3\{2\}$ for all rapidities, with a slight underestimate of $v_2$ at mid-rapidity for the most peripheral events and a slight overestimate for all rapidities for the most central events.

\begin{figure}[ht]
  \hspace{-1.3cm}
  \begin{minipage}[b]{0.55\linewidth}
    \centering
    \includegraphics[width=1.1\textwidth]{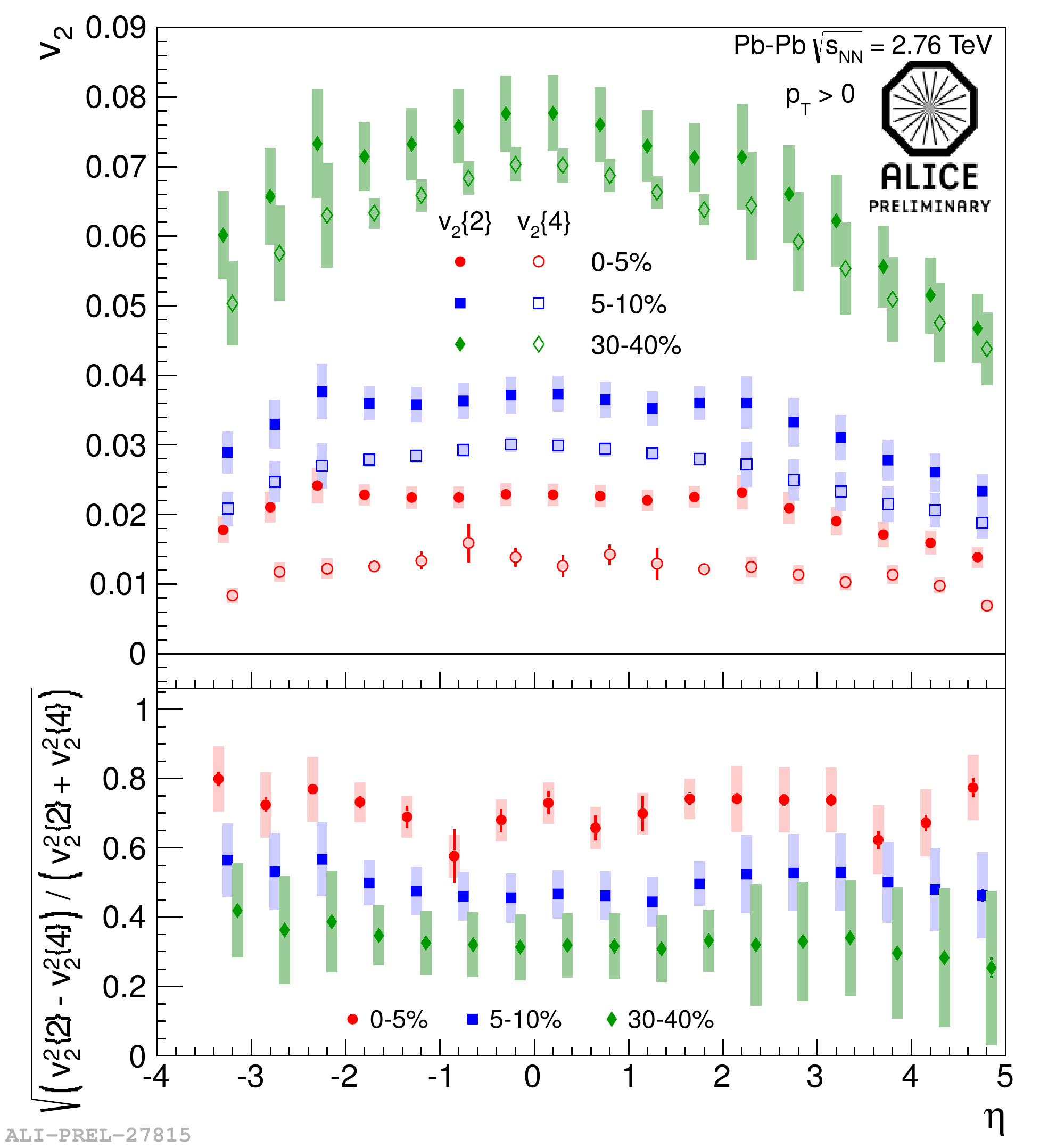}
    \caption{(color online) $v_2\{2\}$ and $v_2\{4\}$ vs. $\eta$ and elliptic flow fluctuations, lines show the statistical uncertainty, bands the systematic uncertainty.}
    \label{f.fluca}
  \end{minipage}
  \hspace{0.50cm}
  \begin{minipage}[b]{0.55\linewidth}
    \centering
    \includegraphics[width=1.1\textwidth]{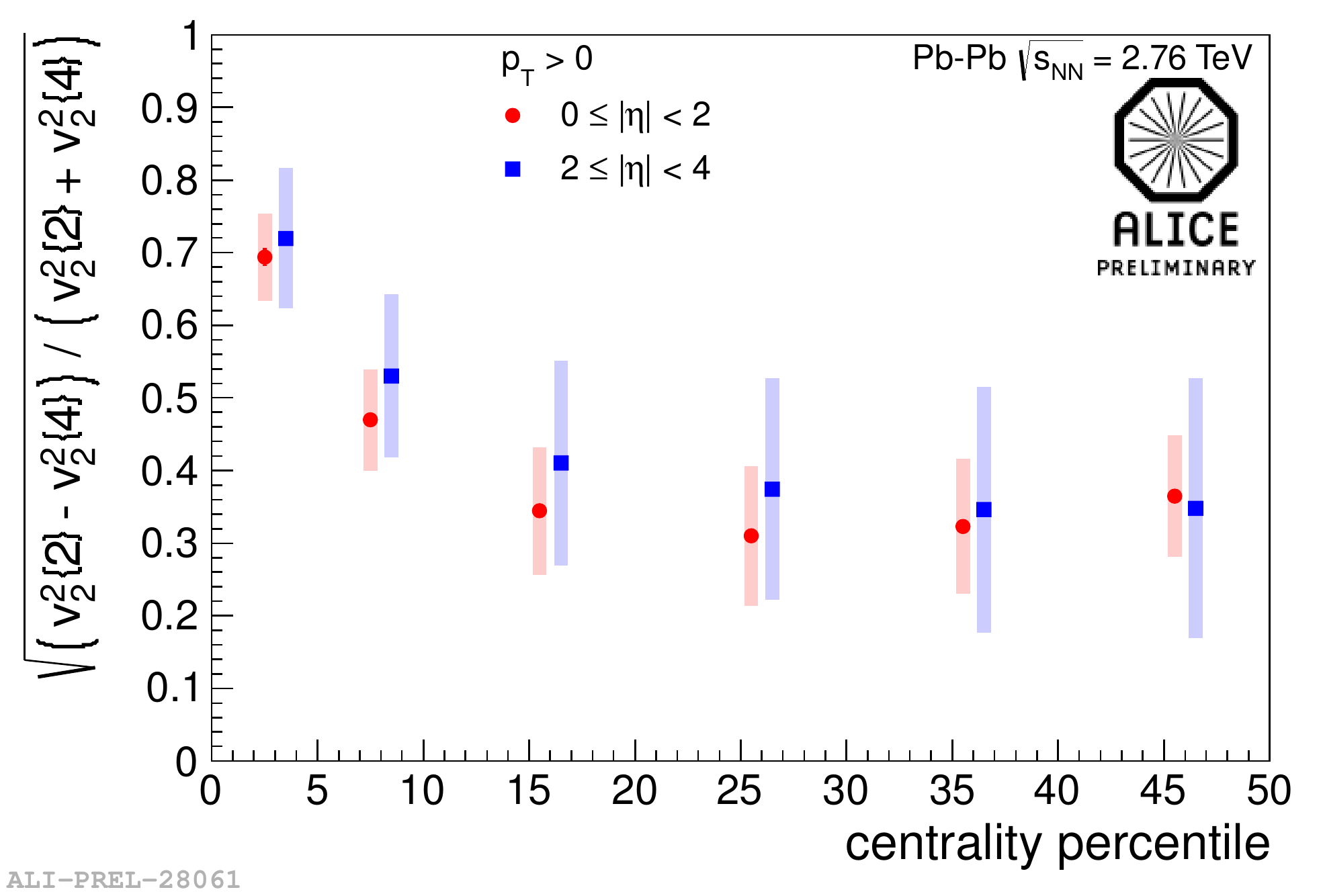}
    \caption{(color online) Elliptic flow fluctuations vs. centrality at mid-rapidity and forward-rapidity.}
    \label{f.flucb}
    \vspace{1cm}
  \end{minipage}
\end{figure}

A comparison of $v_2\{2\}$ and $v_2\{4\}$ is shown in Fig.~\ref{f.fluca}, the shift caused by fluctuations is clearly seen. In the bottom panel the flow fluctuations are estimated, they are found to have a strong centrality dependence, but within the errors no rapidity dependence is observed. 
The centrality dependence of the flow fluctuations (Fig.~\ref{f.flucb}) is observed to be strongest for the most central events. This is consistent with earlier results at mid-rapidity \cite{Collaboration:2011yba}, here we show the same for all $|\eta| < 5$.

\section{Summary}
We have reported the results on the rapidity dependence of elliptic and triangular flow. Elliptic flow was found to have a strong centrality dependence at all rapidities, while triangular flow was found to depend weakly on centrality. Comparing with RHIC measurements, elliptic flow was found to exhibit longitudinal scaling up to LHC energies. The longitudinal scaling is now observed over two orders of magnitude in collision energy.

The AMPT model with parameters tuned to mid-rapidity were found to give a reasonable description for all rapidities, except in the most central and most peripheral, where it overestimates and underestimates the elliptic flow respectively.

The elliptic flow fluctuations were also studied, and were found to be independent of rapidity within the uncertainties. The centrality dependence for all rapidities was found to be consistent with results previously reported for mid-rapidity, with the central events having the largest flow fluctuations. 

%


\end{document}